\documentclass[aps,apl,twocolumn,showpacs]{revtex4-1}
\usepackage{graphicx}
\usepackage{dcolumn}
\usepackage{bm}
\usepackage{amsmath}
\usepackage{tabularx}
\begin{document}
\title{  Correlation of exchange bias with magneto-structural effects across the compensation temperature of Co(Cr$_{1-x}$Fe$_{x}$)$_2$O$_4$ (x = 0.05 and 0.075)}
\author{Ram Kumar$^1$, R. Padam$^1$, S. Rayaprol$^2$, V. Siruguri$^2$ and D.Pal$^{1,}$}
\affiliation{Department of Physics, Indian Institute of Technology
Guwahati, Guwahati, Assam 781039, India}
\affiliation{$^2$UGC-DAE Consortium for Scientific Research-Mumbai Center, R5 Shed, BARC Campus, Trombay, Mumbai 400085, India}

\date{\today}

\begin{abstract}
A small amount of Fe (5$\%$ and 7.5$\%$) substitution in the Cr-site of the multiferroic compound CoCr$_2$O$_4$ leads to a magnetization reversal. In these compounds, we report a sign change in the exchange bias across the compensation temperature, accompanied by a non-monotonic change in the local moments across the compensation temperature. Such non-monotonic change in the magnetic moments is triggered by a similar change in the lattice structure. We relate here the sign change of exchange bias with that of the crystalline energy of the lattice and the Zeeman energy term arising from the anti-site disorder.
\end{abstract}

\maketitle
\section{Introduction}

Cobalt chromite (CoCr$_2$O$_4$) is a well known multiferroic material having fascinating temperature and magnetic field dependent magnetoelectric (ME) properties[1]. In this compound, which has spinel structure, the Co$^{2+}$$-$ ions (3d$^{6}, S$=$ 2$) occupy the tetrahedral sites and the Cr$^{3+}$$-$ ions (3d$^{3}$, S$=$ 3$/$2) are at the octahedral sites. The strong coupling between the magnetic ions in different sites leads to a complex magnetic phase diagram for  CoCr$_2$O$_4$[1]. Eventhough the compound shows a ferrimagnetic ordering below \emph{T$_c$ $\sim$}94$-$97K, we do not observe a sign change in the magnetization across the compensation temperature, as seen in most of the traditional ferrimagnetic compounds. Also, it is reported that the magnetic structure changes below 26K which may be hindering the magnetization reversal[2].

Recently, exchange bias (EB) in materials is considered as one of the very important functionalities. However, the origin of this exchange bias is mostly discussed in terms of a FM and AFM interface. Recently, some homogeneous rare-earth intermetallic compounds have also demonstrated exchange bias across the compensation temperature in single phase compounds[4-6]. Such investigations were extended to compensated ferrimagnetic Heusler alloys where a similar effect was observed [7]. In all the above alloys, the EB is observed because of the presence of the ferromagnetic clusters created due to anti-site disorder in the fully compensated host. However, in  Heusler alloys[7], no reversal of the sign of EB was observed.

This prompted us to substitute Fe in CoCr$_2$O$_4$ so that magnetic compensation could be induced. In this paper, we report EB across this compensation temperature, which possibly arises because of ferromagnetic clusters present in the compensated host. We further report from neutron diffraction measurements that there is a change in the sign of the total magnetic moment across the compensation temperature which, in turn, induces a change in the sign of the exchange bias.

\section{Experimental Details}

Polycrystalline samples of Co(Cr$_{1-x}$Fe$_{x}$)$_2$O$_4$ (x = 0.05 and 0.075) were prepared by standard solid state reaction method  by taking stoichiometric mixtures of the constituent oxides (Co$_3$O$_4$, Cr$_2$O$_3$, and Fe$_2$O$_3$). Neutron diffraction (ND) experiments were carried out on finely ground powders of Co(Cr$_{1-x}$Fe$_{x}$)$_2$O$_4$ (x = 0.05 and 0.075) samples packed in a vanadium container and loaded into a closed cycle cryostat (AS Scientific, UK) for measuring temperature dependent ND patterns at the UGC-DAE CSR beam line at Dhruva reactor using a focusing crystal based powder diffractometer (FCD$-$PD3) at a wavelength of 1.48\emph{$\AA$} [8]. The samples are first cooled in zero field from 90 to 30 K, and ND patterns were recorded in the vicinity of the compensation point at selected temperatures between 30 and 90 K in warming cycle in zero field. Magnetization measurements were carried out as a function of field (up to a maximum field of 7T) in the temperature range 3$-$180 K in a SQUID-vibrating sample magnetometer (Squid-VSM, Quantum Design Inc., USA).

\section{RESULTS AND DISCUSSIONS}
\begin{figure}[!h]
\centering
\includegraphics[width=8cm,height=12cm]{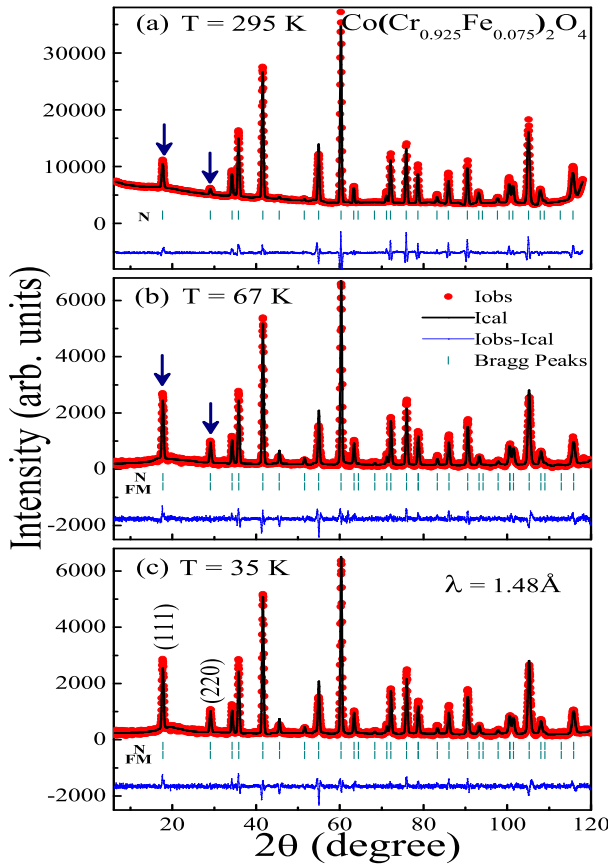}
\caption{\label{refinement data}  Rietveld refinement of ND patterns recorded in zero field at \emph{T}=295, 67, 35 K respectively, are shown here. The patterns at 35 K and 67 K are refined using a structural phase (indicated by the first row of vertical tick marks), and the FM phase (second row), whereas at 295 K only the structural (nuclear) phase is refined. Arrows in the panels (a) and (b) indicate the (111) and (220) peaks.}
\end{figure}

\begin{figure}[b]
\centering
\includegraphics[width=8cm,height=12cm]{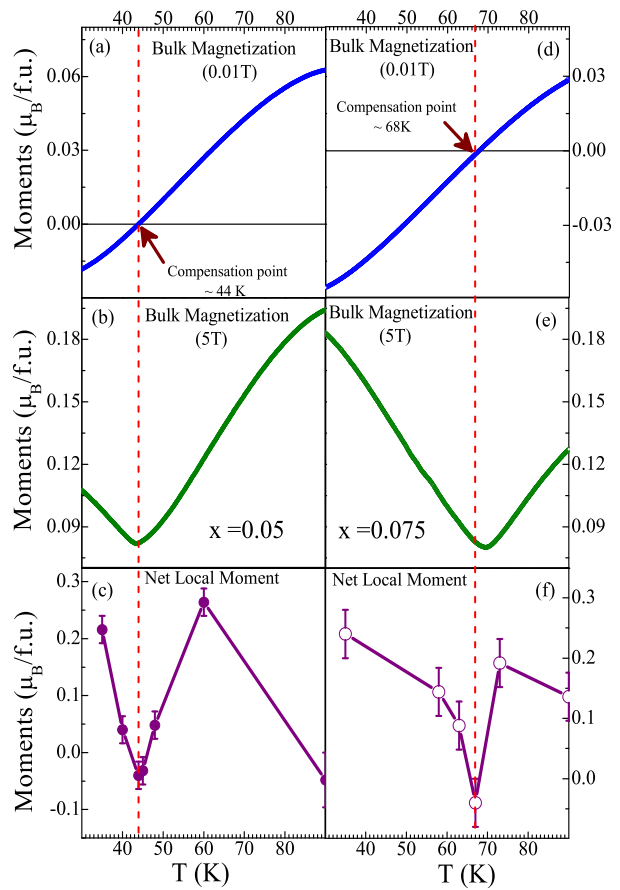}
\caption{\label{combo MvsT 5T 0.01T and ND} Temperature dependent magnetization M(T) curves of Co(Cr$_{1-x}$Fe$_{x}$)$_2$O$_4$ (x = 0.05, left column and 0.075, right column) samples measured under low field 0.01T (a)-(d), high field 5T (b)-(e). (c) and (f) are the net local moments obtained from ND data.}
\end{figure}

\begin{figure}[h]
\centering
\includegraphics[width=8cm,height=12cm]{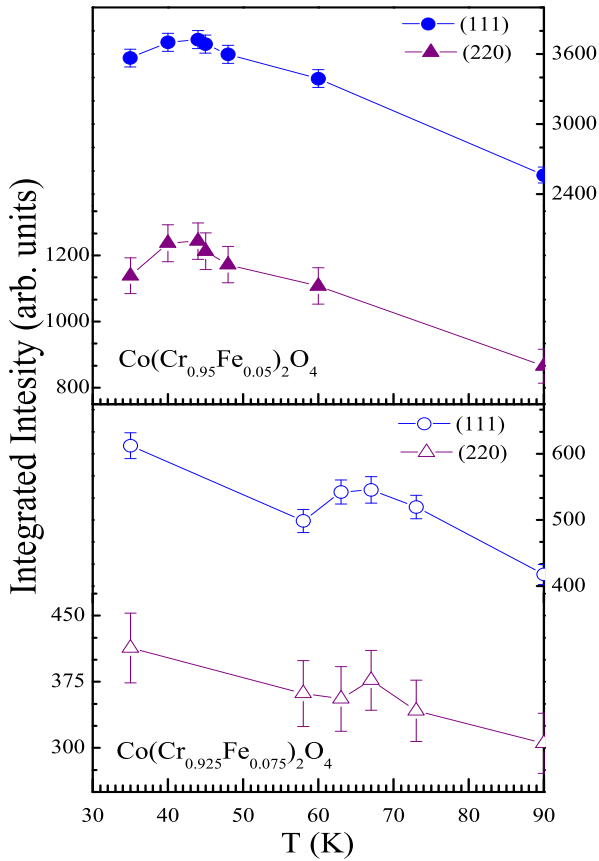}
\caption{\label{combo Integrated Intensity of 111 and 220 peak}  Temperature dependence of the integrated intensity of the (111) and (220) fundamental reflections for both Co(Cr$_{0.095}$Fe$_{0.05}$)$_2$O$_4$ (upper panel) and Co(Cr$_{0.925}$Fe$_{0.075}$)$_2$O$_4$ (lower panel) compounds.}
\end{figure}

\begin{figure}
\centering
\includegraphics[width=9cm,height=12cm]{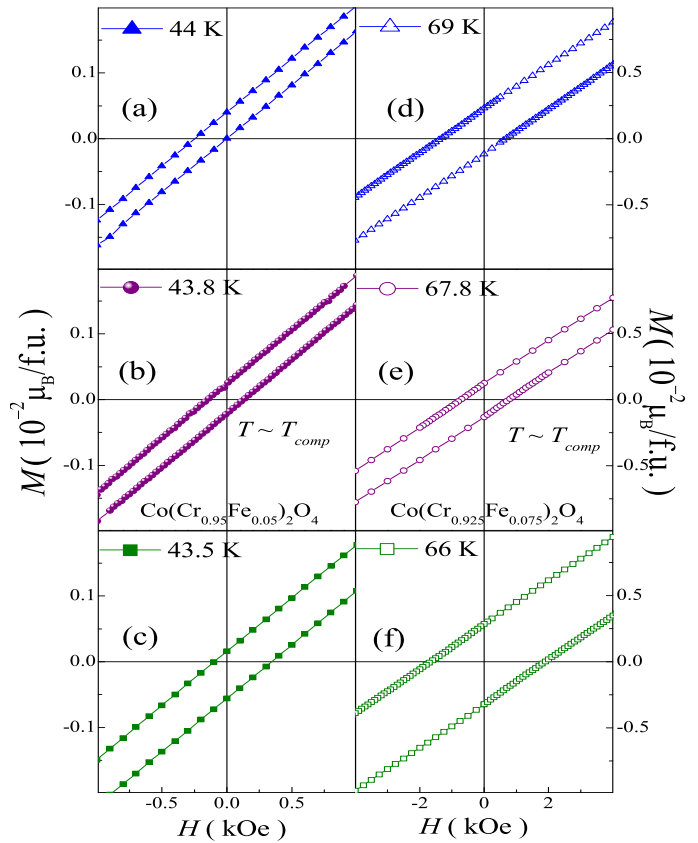}
\caption{\label{EB}  Enlarged view of \emph{M-H} loops of Co(Cr$_{1-x}$Fe$_{x}$)$_2$O$_4$ (x = 0.05 and 0.075) samples in the vicinity of \emph{T$_{comp}$$\approx$}43.8 K and \emph{T$_{comp}$$\approx$}67.8 K for x = 0.05 and 0.075 samples, respectively.}
\end{figure}

\begin{figure*}[!tbp]
\centering
\begin{minipage}[b]{0.45\textwidth}
\includegraphics[trim=0cm 0cm 0.15cm 0cm, clip=true, scale=0.50]{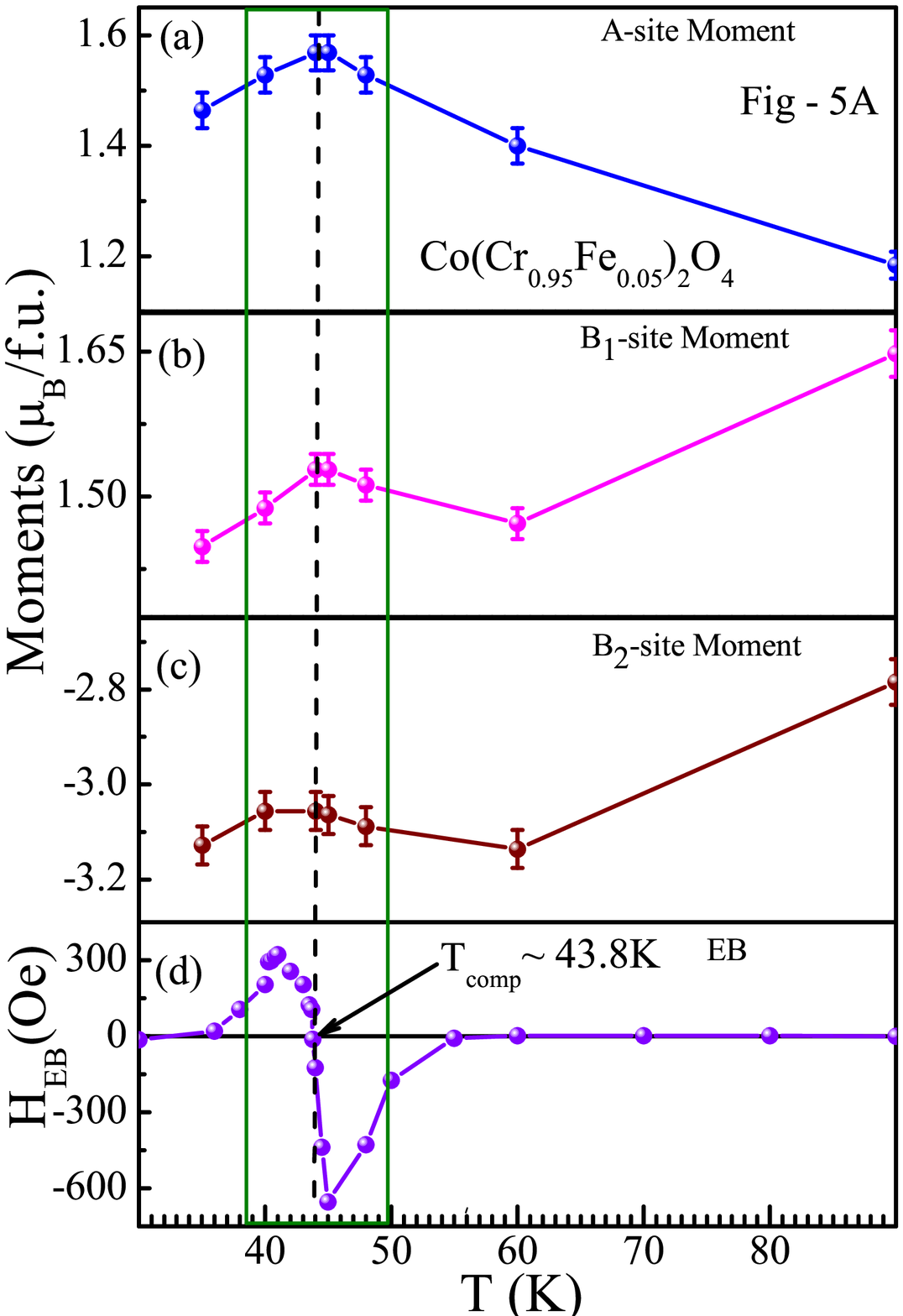}
\end{minipage}
\hfill
\begin{minipage}[b]{0.45\textwidth}
\includegraphics[trim=0.25cm 0cm 0cm 0cm, clip=true, scale=0.50]{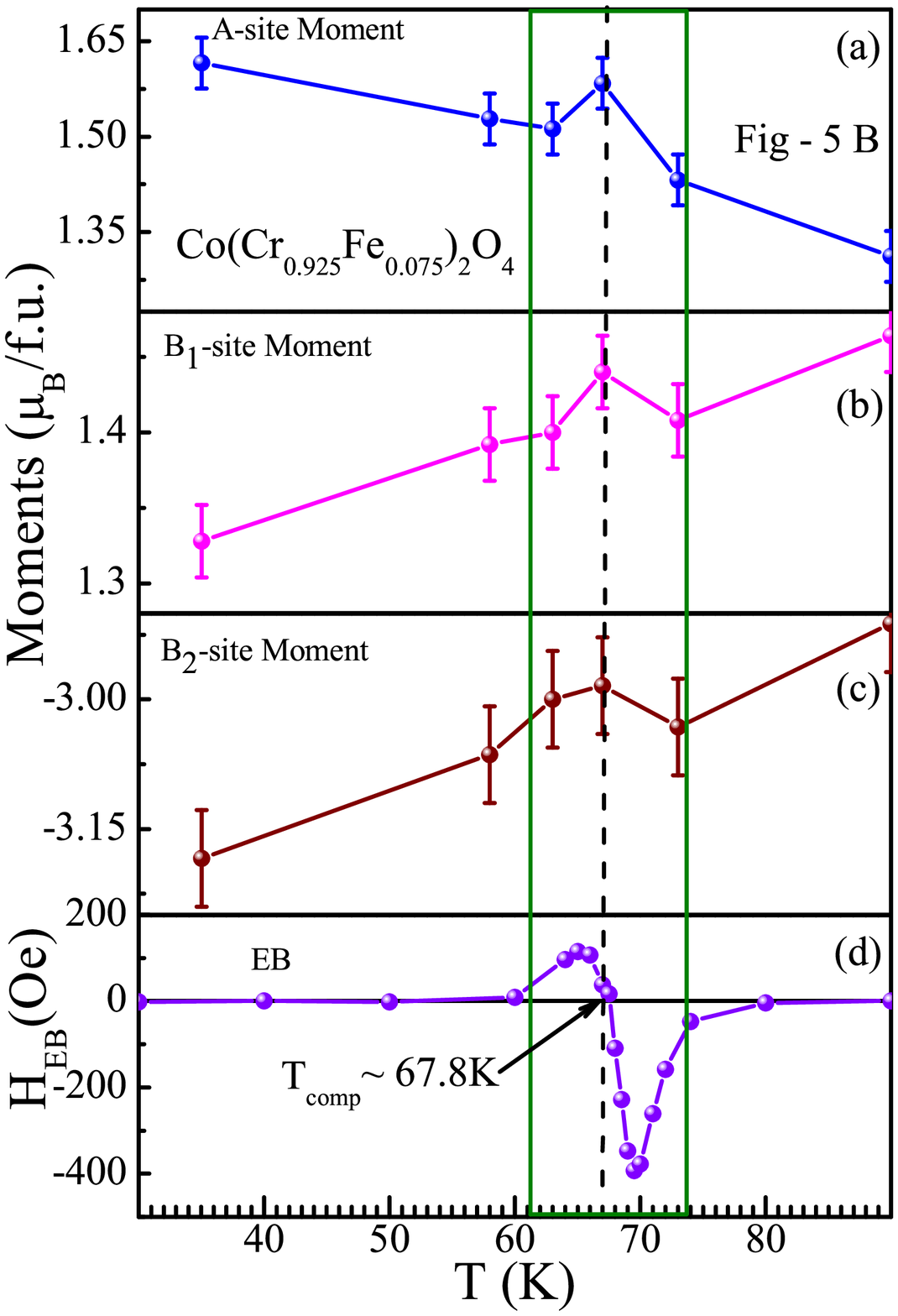}
\end{minipage}
\hfill
\caption{\label{Lattce AB1B2momEB combo}  Temperature dependence of different site moments and exchange bias field of Co(Cr$_{0.095}$Fe$_{0.05}$)$_2$O$_4$ (Fig- 5A) and Co(Cr$_{0.925}$Fe$_{0.075}$)$_2$O$_4$ (Fig- 5B) compounds.}
\end{figure*}

It is known that EB phenomenon is an interfacial property. Hence, it is necessary to investigate the phase purity thoroughly when it is observed in single phase materials. We recorded ND patterns in the paramagnetic state to determine the crystal structure. Fig.1(a) shows the Rietveld refinement of ND pattern of 7.5$\%$ Fe substituted sample recorded at 295 K using the FULLPROF program[9]. We can observe that the experimental data can be fitted well to the cubic spinel structure with \emph{Fd}$\overline{3}$\emph{m} space group, without any impurities. Co atoms occupy Wyckoff position 8\emph{a}(1/8, 1/8, 1/8), Cr and Fe atoms occupy the position 16\emph{d}(1/2, 1/2, 1/2) and Oxygen atoms occupy the position 32\emph{e}( 0.26, 0.26, 0.26). Lattice parameters of 5$\%$ and 7.5$\%$ Fe samples at room temperature are found to be 8.3381$\AA$ and 8.3399$\AA$, respectively. A small expansion in the unit cell parameter with Fe substitution is seen. Fig.1(b) and (c) show the ND patterns in the magnetically ordered region of 7.5$\%$ Fe sample (i.e., below the ordering temperature \emph{T$_c$  $\sim$} 118K). Due to the ferrimagnetic ordering of moments below \emph{T$_c$}, we can see a clear increase of some nuclear Bragg peaks pointing to the appearance of  ferromagnetic order with propagation vector \emph{$\kappa$} =[0 0 0]. In order to follow the temperature dependent behavior of the FM components, Rietveld refinement was carried out for data in the temperature range of 30$-$90K for both the compounds. Hence, in the magnetically ordered state, in addition to nuclear reflections, we have determined the propagation vector \emph{$\kappa$} for magnetic reflections using the program K-Search, which is also a part of the FULLPROF suite program. For ferromagnetic/ferrimagnetic systems, \emph{$\kappa$} is always \emph{$\kappa$} =[0 0 0] since no superstructure reflections are observed. It was found to satisfy the Bragg conditions for all ferromagnetic (FM) peaks at all temperatures above 30K. Below 30K, we have observed an incommensurate phase similar to the parent compound, CoCr$_2$O$_4$ which matches with the result reported by Menyuk et. al[2-3].

Six panels of Fig. 2 depict the comparison of the temperature dependent spontaneous magnetization obtained from the refinement of the magnetic phase of the ND patterns of 5$\%$ and 7.5$\%$ Fe samples with that of the bulk magnetization measured experimentally under the applied fields of 0.01T and 5T. We can see that, similar to the bulk magnetization, the moments obtained from the ND data also exhibit near zero values at the compensation points \emph{T$_{comp}$ $\approx$} 43.8K and \emph{$\approx$} 67.8K for 5$\%$ and 7.5$\%$ Fe (See Fig. 2(a) and (d)), respectively. And remarkably, it can be observed that the behavior of the spontaneous magnetization across the compensation point is very similar to the bulk magnetization under 5T field. One more point that is clear from these combined plots is that magnetization obtained from both measurements is of the same order for the compounds. From this, it is understood that there are orientations of the moments corresponding to different sub-lattices, across the compensation point. All these features are not noticed in case of the parent CoCr$_2$O$_4$ compound.

Further, Fig.3 presents the temperature dependent variation of integrated intensity of the fundamental reflections[2] (111) and (220) of 5$\%$ and 7.5$\%$ Fe substituted samples calculated from ND data. Intensity of both the reflections appears to increase up to 30K but there is an unusual change around compensation temperature \emph{T$_{comp}$ $\approx$} 43.8K and \emph{$\approx$} 67.8K for 5$\%$ and 7.5$\%$ Fe samples, respectively. It was observed in the parent compound, CoCr$_2$O$_4$, the magnetic order consists of a ferrimagnetic component and a spiral component below the ferrimagnetic transition temperature. Hence, in this case it was found that the magnetic moment makes an angle (cf. cone angle) with the cone axis in the [0 0 1] direction [2,11]. In our compounds, where the amount of Fe substitution is minimal, the observed behavior is similar to that of parent compound. The variation in the magnitude of these spiral components is the primary cause for the variation in the integrated intensity. Hence, this unusual change of integrated intensity around compensation temperature could be the result of change in the magnitude of the spiral components (i.e. the cone angles) of the magnetic moments.

A set of six panels in Fig. 4 shows the distinct and detailed information on the hysteretic behavior of field cooled \emph{M-H} loops for 5$\%$ and 7.5$\%$ Fe compounds in the vicinity of the \emph{T$_{comp}$ $\approx$} 43.8K and \emph{$\approx$} 67.8K. The small left shift in the origin of \emph{M-H} loop observed above the \emph{T$_{comp}$} of respective sample becomes right shift below \emph{T$_{comp}$}. In Fig. 5, we show the temperature dependence of \emph{EB} along with the moments corresponding to different sub-lattices in the spinel structure. \emph{EB} field defined as, \emph{H$_{EB}$ = (H$_+$ + H$_-$)}/2, where \emph{H$_+$} and \emph{H$_-$} represent the right and left field values of \emph{M-H} loop where the net magnetization crosses the \emph{M=0} axis. It can be seen that \emph{EB} field changes its sign across the \emph{T$_{comp}$} for both the samples. Curiously, in the narrow temperature window in the proximity of \emph{T$_{comp}$}, there is a non-monotonic variation in the individual moments of A-site, B$_1$-site and B$_2$-site. This is  indicative of the reorientation of moments due to the underlying spin-reorientation which was invoked for the explanation of the sign reversal of EB across the compensation temperature.\\
In the parent compound CoCr$_2$O$_4$ below \emph{T$_c$ $\sim$}94$-$97K, the spin configurations of the sub-lattices A, B$_1$ and B$_2$ take a complicated conical spin spiral form. The propagation vector of these spin spiral configurations of all these sub-lattices points to the [1 1 0] direction and the angle made by these spins with the [0 0 1] direction is defined as the cone angle for various sub-lattices. This compound has a strong coupling between the spin configuration and the crystalline lattice. So, with the change of temperature, there is a change in the lattice and this should result in a change in the cone angles of the spin configurations of all the sub-lattices[2,12]. Hence, it is interesting to observe a variation of the cone angles across the compensation temperature in these Fe substituted compounds. Fig. 6(a), (b) and (c) depict the change in the cone angles($\alpha$, $\beta$ and $\gamma$) of different sites(A, B$_1$ and B$_2$) as a function of temperature for x=0.075 compound. The panels (a) and (b) show that there is a sudden dip in the cone angle at the compensation temperature. Similarly, the panel (c) shows an increase in the cone angle across the compensation temperature. The panel (d) shows the variation of lattice constant,'a' as a function of temperature. We observe a slight increase of the lattice constant below the \emph{T$_{comp}$} (67.8K) till 58K. This unusual variation of the lattice constant results in the non-monotonic changes in the cone angles. We observe a similar pattern of plots for the other compound (i.e, x =0.05).

\begin{figure}
\centering
\includegraphics[width=9cm,height=9cm]{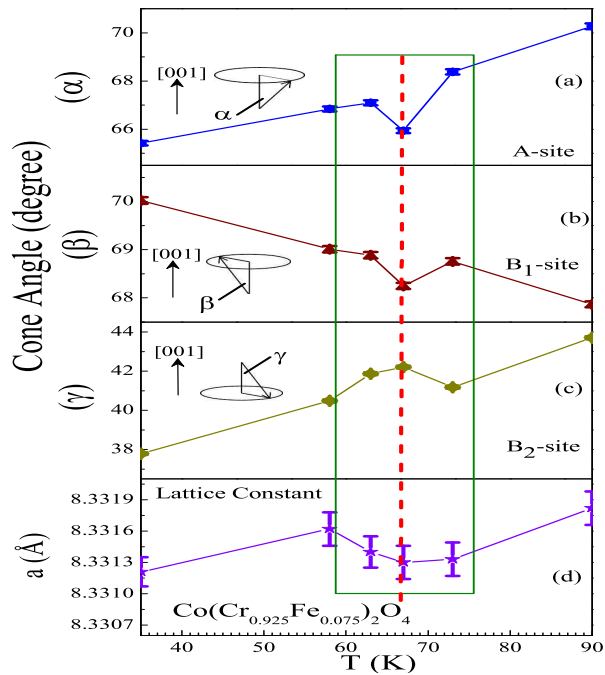}
\caption{\label{Cone angle} Temperature dependence  of (a) cone angle ($\alpha$) of the sub-lattice A, (b) cone angle ($\beta$) of the sub-lattice B$_1$, (c) cone angle ($\gamma$) of the sub-lattice B$_2$ and (d) Lattice constant as of 7.5$\%$ Fe sample. Inset shows the ferrimagnetic spin spiral order which points to the [001] direction similar to that of the CoCr$_2$O$_4$[2,11].}
\end{figure}

Recently, it has been observed that exchange bias can be generated in a fully compensated ferrimagnetic Heusler alloys having intrinsic anti-site disorder[7]. These anti-site disorders give rise to small ferromagnetic clusters in the antiferromagnetic host, which also results in a small lack of compensation. The exchange interaction between the ferromagnetic clusters and antiferromagnetic host gives rise to this exchange bias which can be of the order of a few tesla. In these Fe substituted CoCr$_2$O$_4$ compounds, we also observe similar exchange bias effect close to the compensation temperatures (\emph{$\approx$}43.8K and \emph{$\approx$}67.8K). This may be due to the presence of small ferromagnetic clusters arising due to anti-site disorder in the AFM host. Normal exchange bias effect, which is negative  arises due to the exchange interaction between the FM cluster and host material. This is what we observe (cf. Fig. 5) above the compensation temperature in both the compounds. However, the positive EB which we observe below compensation temperature is unusual. In the literature, such an effect is discussed in context of the crystalline energy[13-15]. Fig. 6(d) shows that the lattice constant increases with the decrease in temperature (67.8$ - $58K) below the compensation temperature, whereas in the rest of the temperature range, it is observed to monotonically decrease with the decrease of temperature. Below the compensation temperature, there is a dominance of the crystalline energy over the Zeeman energy, resulting in the positive exchange bias in this temperature window. Such dominance of the crystalline energy is also reflected in the rotation of the cone angles of the local magnetic moments of different sites(cf. Fig. 6 (a-c)).

\section{Conclusions}
We have reported here neutron diffraction experiments on Co(Cr$_{1-x}$Fe$_{x}$)$_2$O$_4$ (x = 0.05 and 0.075) compounds at low temperature in the vicinity of compensation temperature. The local magnetic moments and the crystalline structure show a non monotonic variation across the compensation temperature. We also observed a sign change in exchange bias across this compensation temperature and correlated this with the magneto-structural effects revealed from the neutron data. This class of materials provides a platform to study the competition between the crystalline energy and the Zeeman energy, which is reflected in the sign change of the exchange bias. Such effects in these homogeneous compounds call for further exploration in the form of single crystal studies of these compounds and of other compensated compounds, where the lattice and the spin structure are strongly coupled.\\
\section{Acknowledgments}
D.P. acknowledge the UGC-DAE Consortium for Scientific Research, Mumbai Centre, India for support in the form of a collaborative research project (Project No. CRS-M-195). D.P. and R.K. acknowledge FIST programme of Department of Science and Technology, India for partial support of this work.

\end{document}